\documentclass[twocolumn,preprintnumbers,superscriptaddress,showpacs,amssymb,amsmath]{revtex4} 

\begin{document}

\title{{\large Differential cross section of $\gamma n \rightarrow K^+ \Sigma^- $ on bound neutrons with 
incident photons from 1.1 to 3.6 GeV}}

\newcommand*{\INFNFR}{INFN, Laboratori Nazionali di Frascati, 00044 Frascati, Italy}
\newcommand*{\INFNFRindex}{1}
\affiliation{\INFNFR}
\newcommand*{\JMU}{James Madison University, Harrisonburg, Virginia 22807}
\newcommand*{\JMUindex}{2}
\affiliation{\JMU}
\newcommand*{\JLAB}{Thomas Jefferson National Accelerator Facility, Newport News, Virginia 23606}
\newcommand*{\JLABindex}{3}
\affiliation{\JLAB}
\newcommand*{\ANL}{Argonne National Laboratory, Argonne, Illinois 60441}
\newcommand*{\ANLindex}{4}
\affiliation{\ANL}
\newcommand*{\ASU}{Arizona State University, Tempe, Arizona 85287-1504}
\newcommand*{\ASUindex}{5}
\affiliation{\ASU}
\newcommand*{\UCLA}{University of California at Los Angeles, Los Angeles, California  90095-1547}
\newcommand*{\UCLAindex}{6}
\affiliation{\UCLA}
\newcommand*{\CSUDH}{California State University, Dominguez Hills, Carson, CA 90747}
\newcommand*{\CSUDHindex}{7}
\affiliation{\CSUDH}
\newcommand*{\CANISIUS}{Canisius College, Buffalo, NY}
\newcommand*{\CANISIUSindex}{8}
\affiliation{\CANISIUS}
\newcommand*{\CMU}{Carnegie Mellon University, Pittsburgh, Pennsylvania 15213}
\newcommand*{\CMUindex}{9}
\affiliation{\CMU}
\newcommand*{\CUA}{Catholic University of America, Washington, D.C. 20064}
\newcommand*{\CUAindex}{10}
\affiliation{\CUA}
\newcommand*{\SACLAY}{CEA, Centre de Saclay, Irfu/Service de Physique Nucl\'eaire, 91191 Gif-sur-Yvette, France}
\newcommand*{\SACLAYindex}{11}
\affiliation{\SACLAY}
\newcommand*{\CNU}{Christopher Newport University, Newport News, Virginia 23606}
\newcommand*{\CNUindex}{12}
\affiliation{\CNU}
\newcommand*{\UCONN}{University of Connecticut, Storrs, Connecticut 06269}
\newcommand*{\UCONNindex}{13}
\affiliation{\UCONN}
\newcommand*{\ECOSSEE}{Edinburgh University, Edinburgh EH9 3JZ, United Kingdom}
\newcommand*{\ECOSSEEindex}{14}
\affiliation{\ECOSSEE}
\newcommand*{\FU}{Fairfield University, Fairfield CT 06824}
\newcommand*{\FUindex}{15}
\affiliation{\FU}
\newcommand*{\FIU}{Florida International University, Miami, Florida 33199}
\newcommand*{\FIUindex}{16}
\affiliation{\FIU}
\newcommand*{\FSU}{Florida State University, Tallahassee, Florida 32306}
\newcommand*{\FSUindex}{17}
\affiliation{\FSU}
\newcommand*{\GWU}{The George Washington University, Washington, DC 20052}
\newcommand*{\GWUindex}{18}
\affiliation{\GWU}
\newcommand*{\ISU}{Idaho State University, Pocatello, Idaho 83209}
\newcommand*{\ISUindex}{19}
\affiliation{\ISU}
\newcommand*{\INFNGE}{INFN, Sezione di Genova, 16146 Genova, Italy}
\newcommand*{\INFNGEindex}{20}
\affiliation{\INFNGE}
\newcommand*{\INFNRO}{INFN, Sezione di Roma Tor Vergata, 00133 Rome, Italy}
\newcommand*{\INFNROindex}{21}
\affiliation{\INFNRO}
\newcommand*{\ORSAY}{Institut de Physique Nucl\'eaire ORSAY, Orsay, France}
\newcommand*{\ORSAYindex}{22}
\affiliation{\ORSAY}
\newcommand*{\ITEP}{Institute of Theoretical and Experimental Physics, Moscow, 117259, Russia}
\newcommand*{\ITEPindex}{23}
\affiliation{\ITEP}
\newcommand*{\KNU}{Kyungpook National University, Daegu 702-701, Republic of Korea}
\newcommand*{\KNUindex}{24}
\affiliation{\KNU}
\newcommand*{\LPSC}{LPSC, Universit\'e Joseph Fourier, CNRS/IN2P3, INP, Grenoble, France}
\newcommand*{\LPSCindex}{25}
\affiliation{\LPSC}
\newcommand*{\UNH}{University of New Hampshire, Durham, New Hampshire 03824-3568}
\newcommand*{\UNHindex}{26}
\affiliation{\UNH}
\newcommand*{\NSU}{Norfolk State University, Norfolk, Virginia 23504}
\newcommand*{\NSUindex}{27}
\affiliation{\NSU}
\newcommand*{\OHIOU}{Ohio University, Athens, Ohio  45701}
\newcommand*{\OHIOUindex}{28}
\affiliation{\OHIOU}
\newcommand*{\ODU}{Old Dominion University, Norfolk, Virginia 23529}
\newcommand*{\ODUindex}{29}
\affiliation{\ODU}
\newcommand*{\RPI}{Rensselaer Polytechnic Institute, Troy, New York 12180-3590}
\newcommand*{\RPIindex}{30}
\affiliation{\RPI}
\newcommand*{\URICH}{University of Richmond, Richmond, Virginia 23173}
\newcommand*{\URICHindex}{31}
\affiliation{\URICH}
\newcommand*{\ROMAII}{Universita' di Roma Tor Vergata, 00133 Rome Italy}
\newcommand*{\ROMAIIindex}{32}
\affiliation{\ROMAII}
\newcommand*{\MSU}{Skobeltsyn Nuclear Physics Institute, 119899 Moscow, Russia}
\newcommand*{\MSUindex}{33}
\affiliation{\MSU}
\newcommand*{\SCAROLINA}{University of South Carolina, Columbia, South Carolina 29208}
\newcommand*{\SCAROLINAindex}{34}
\affiliation{\SCAROLINA}
\newcommand*{\UNIONC}{Union College, Schenectady, NY 12308}
\newcommand*{\UNIONCindex}{35}
\affiliation{\UNIONC}
\newcommand*{\UTFSM}{Universidad T\'{e}cnica Federico Santa Mar\'{i}a, Casilla 110-V Valpara\'{i}so, Chile}
\newcommand*{\UTFSMindex}{36}
\affiliation{\UTFSM}
\newcommand*{\ECOSSEG}{University of Glasgow, Glasgow G12 8QQ, United Kingdom}
\newcommand*{\ECOSSEGindex}{37}
\affiliation{\ECOSSEG}
\newcommand*{\VIRGINIA}{University of Virginia, Charlottesville, Virginia 22901}
\newcommand*{\VIRGINIAindex}{38}
\affiliation{\VIRGINIA}
\newcommand*{\WM}{College of William and Mary, Williamsburg, Virginia 23187-8795}
\newcommand*{\WMindex}{39}
\affiliation{\WM}
\newcommand*{\YEREVAN}{Yerevan Physics Institute, 375036 Yerevan, Armenia}
\newcommand*{\YEREVANindex}{40}
\affiliation{\YEREVAN}
 
\newcommand*{\NOWCUA}{Catholic University of America, Washington, D.C. 20064}
\newcommand*{\NOWJLAB}{Thomas Jefferson National Accelerator Facility, Newport News, Virginia 23606}
\newcommand*{\NOWCNU}{Christopher Newport University, Newport News, Virginia 23606}
\newcommand*{\NOWECOSSEE}{Edinburgh University, Edinburgh EH9 3JZ, United Kingdom}
\newcommand*{\NOWWM}{College of William and Mary, Williamsburg, Virginia 23187-8795}
\newcommand*{\NOWLANL}{Los Alamos National Laboratory, Los Alamos, New Mexico 87545}

\author {S.~Anefalos~Pereira} 
\affiliation{\INFNFR}
\author {M.~Mirazita} 
\affiliation{\INFNFR}
\author {P.~Rossi} 
\affiliation{\INFNFR}
\author {E.~De~Sanctis} 
\affiliation{\INFNFR}
\author {G.~Niculescu} 
\affiliation{\JMU}
\author {I.~Niculescu} 
\affiliation{\JMU}
\author {S.~Stepanyan} 
\affiliation{\JLAB}
\author {K.P.~Adhikari} 
\affiliation{\ODU}
\author {M.~Aghasyan} 
\affiliation{\INFNFR}
\author {M.~Anghinolfi} 
\affiliation{\INFNGE}
% ...

\collaboration{The CLAS Collaboration}
\noaffiliation

\date{\today}

\begin{abstract}
Differential cross sections of the reaction $\gamma d \rightarrow K^+ \Sigma^- (p)$ have been measured 
with the CLAS detector at Jefferson Lab using incident photons with energies between 1.1 and 3.6 GeV.
This is the first complete set of strangeness photoproduction data on the neutron covering a broad angular 
range. At energies close to threshold and up to $E_{\gamma} \sim$ 1.8 GeV,
 the shape of the angular distribution is suggestive of the presence of $s$-channel production mechanisms.
For $E_{\gamma} > 1.8$~GeV, a clear forward peak appears and becomes more prominent as the 
photon energy increases, suggesting contributions from $t$-channel production mechanisms. 
These data can be used to constrain future analysis of this reaction.
\end{abstract}


\maketitle

\indent
A major goal of hadron physics is to study the structure of the nucleon and its excited 
states. However, understanding nucleon resonance excitation is a serious challenge due to the 
non-perturbative nature of QCD at low energies. This makes the situation for the excited states of the
nucleon ($N$ and $\Delta$ resonances) still unclear: many more states are predicted than observed and 
states with certain quantum numbers appear at energies much lower than predicted. This has been known 
for a long time as the ``missing resonance'' problem~\cite{isgurekarl}. In quark models 
(see Ref.~\cite{capstick} for reviews), the number of 
excited states is determined by the effective degrees of freedom, while their ordering and
decay properties are related to the residual quark-quark interaction.

The effective degrees of freedom 
in standard non-relativistic quark models are three equivalent valence quarks with one-gluon exchange 
interactions. A different class of models uses interactions that give 
rise to a quark-diquark clustering of the baryon~\cite{anselmino}. If there is a tightly bound diquark, 
only two degrees of freedom are available at low energies; thus, fewer states are predicted. Furthermore, 
selection rules in the decay pattern may arise from the quantum numbers of a diquark. More states are 
provided by collective models of the baryon, like the algebraic approach~\cite{bijker}, or in the 
framework of flux-tube models~\cite{isgurepaton}, which are motivated by lattice QCD. So far, however, 
the experimentally observed number of states is still far lower than predicted by the quark-diquark models.

Experimentally, most of our present knowledge of baryon resonances comes from reactions involving pions in 
the initial and/or final states. For increasing masses, both the energy overlap of the resonances and meson
production make it more difficult to separate the resonance contributions. A possible explanation for the 
missing resonance problem could be that pionic coupling to the intermediate $N^*$ or $\Delta^*$ states 
is weak and that many of the missing states only become visible in other reaction channels. 
Photoproduction of non-strange resonances detected via decay into strange particles offers two 
benefits: (1) two-body $KY$ (where $Y$ denotes any hyperon) final states are easier to analyze than the 
three-body $\pi$$\pi$$N$ final states that dominate the decays at higher masses resonances; (2) couplings 
of nucleon resonances to $KY$ final states are expected to differ from those to $\pi$$N$ and 
$\pi$$\pi$$N$ final states~\cite{capstickN}. Therefore, looking in the strangeness sector casts a different 
light on the resonance excitation spectrum, and thus, may emphasize resonances not revealed in $\pi$$N$ 
scattering. To date, however, the PDG compilation~\cite{pdg} gives poorly known $K\Lambda$ 
couplings for only five well-established resonances, and no $K\Sigma$ couplings for any resonances.
Mapping out the spectrum of excited states that decay into $KY$ particles is therefore crucial to provide 
a deeper insight into the underlying degrees of freedom of the nucleon and to discriminate among 
different models.

The search for missing resonances requires more than the study of the hadronic mass spectrum. In fact, QCD 
cannot be directly tested against experimental $N^*$ mass spectra without a model for the production 
dynamics~\cite{lee}. Thus, in addition to the $s$-channel contributions, important in the resonance region 
in order to reproduce the invariant mass spectra, the $t$- and $u$-channel meson and baryon exchanges are 
also necessary in the theoretical description. The former are needed in order to describe the diffractive 
part of the production, and $u$-channel diagrams are necessary to describe the back-angle production. Thus, 
measurements that can constrain the phenomenology for these reactions are just as important as finding one 
or more of the missing resonances. 

A large amount of cross-section data of hyperon photoproduction on the proton has been published in recent 
years by the SAPHIR~\cite{SAPHIR}, CB-ELSA/TAPS~\cite{TAPS1,TAPS2}, CLAS~\cite{CLAS,McCracken} 
and LEPS~\cite{LEPSp} collaborations from threshold up to $E_{\gamma} \sim 3.8$~GeV over a wide angular 
range. The polarization of the recoil hyperon has also been measured by CLAS~\cite{CLASpol,McCracken}, 
SAPHIR~\cite{SAPHIR} and GRAAL~\cite{GRAALpol}, while photon beam asymmetries have been measured by 
LEPS~\cite{LEPSasym}. Despite this large body of data, theoretical ambiguities still exist. In fact,
theorists have found conflicting evidence for resonances using isobar models~\cite{maid}, coupled-channel 
~\cite{mosel,ani,juliadiaz} and partial wave analysis~\cite{sar} approaches.

In this situation the necessity of more data and from different channels is evident. In particular, 
for $Y$-photoproduction on the neutron, one can take full advantage of the isospin symmetry, 
adding significant constraints on the $\gamma KNY$ coupling constants \cite{maid2}.
Unfortunately, data of hyperon photoproduction on neutrons are very scarce, with the only available data 
from LEPS~\cite{LEPSn}, covering a limited photon energy range at very forward kaon angles.

In this paper high-precision cross sections of the reaction $\gamma d \rightarrow K^+ \Sigma^- (p)$ 
in a broad kinematic range are presented. The data were acquired using the CLAS detector~\cite{clas} 
housed in Hall B at Jefferson Lab. 
A bremsstrahlung photon beam produced by a 3.776 GeV continuous electron beam hitting a  
$10^{-4}$ radiation-lengths gold foil was used \cite{tagging}. Tagged photons, in the energy 
range from 0.8 to 3.6 GeV, were directed onto a liquid-deuterium target. With an electron beam current 
of $\sim 25$ nA, the photon flux incident on the deuterium target was $\sim 10^8$ $\gamma/$s.

The primary kaon, and the pion and neutron coming from the $\Sigma^-$ decay (with branching fraction 
$b_{\Sigma^-} \sim 100\%$) were detected by CLAS. The low-energy spectator proton was reconstructed using
the missing-mass technique.
Fiducial cuts were applied to both real and Monte-Carlo simulated data in order to exclude regions where 
the detector acceptance was not well understood 
and the regions where the drift chambers or scintillator efficiencies were not well known.
Neutral particles are identified in CLAS as clusters in the electromagnetic calorimeters that are not associated with any charged 
track in the drift chambers. Neutral clusters with $\beta > 0.9$ are then identified as photons, while clusters with 
$\beta <0.9$ are associated with neutrons. 

In order to identify good $\gamma d \rightarrow K^+ \pi^- n X$ candidates, with the missing particle $X$ consistent 
with a spectator proton, we first applied a cut on 
the missing momentum $P_{X} \leq 0.25$ GeV/$c$. The remaining events, integrated over all angles, were divided 
into 100 MeV wide bins in photon energy. In each bin, the missing-mass distribution was used to select events 
consistent with a missing spectator proton. 

\begin{figure}[h]
\vspace{3.5cm}                                                                  
\includegraphics{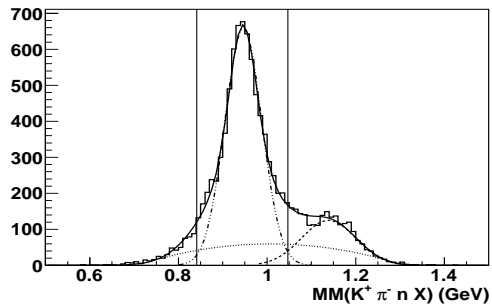}
\caption[]{\small Missing-mass distribution of the $\gamma d \rightarrow K^+ \pi^- n X$ reaction for 
photon energy in the bin 2.0--2.1 GeV. The solid line is the total fit to 
the distribution. The dot-dashed line represents the proton distribution, the dashed line is the 
contribution of the events with an additional $\pi^0$ as explained in the text, and the dotted line 
is the background contribution. The two vertical lines represent a $3\sigma$ cut.}
\label{fig:pro_mass_high}
\end{figure}

The missing-mass distribution for the photon energy bin of 2.0--2.1 GeV in Fig.~\ref{fig:pro_mass_high} 
shows a clear proton peak and a smaller structure at higher masses. The latter, that starts to appear at 
photon energies $\gtrsim 2$ GeV, is due to photoproduction events of $\Sigma^*(1385)^-$ and $K^*(892)^+$. 
The $\Sigma^*(1385)^-$ decays into $\Sigma \pi$ with $b_{\Sigma^*} \sim 12\%$ and the $K^*(892)^+$ decays 
into $K \pi$ with $b_{K^*} \sim 100\%$. Each missing-mass distribution was fit with two Gaussian line 
shapes plus a polynomial curve. The total fit and each contribution separately are shown in 
Fig.~\ref{fig:pro_mass_high}. Events with a spectator proton were selected by applying a $3\sigma$ cut 
around the main peak. The background contribution coming from $\Sigma^*(1385)^-$ and $K^*(892)^+$ events was 
estimated to be between $1$ and $3\%$ and then subtracted.
\\

\begin{figure}[h]
\vspace{3.5cm}                                                                  
\includegraphics{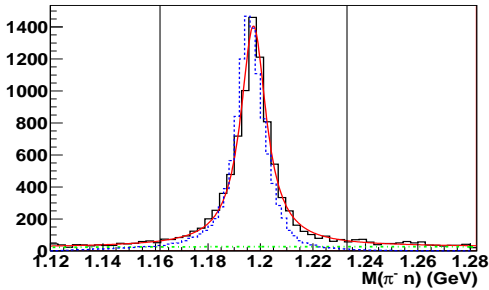}
\caption[] {\small (color online) The $n \pi^-$ invariant mass distribution for data (full histogram) 
and MC (dashed histogram) in the photon energy range 2.0--2.1 GeV. The solid line is the fit of the distribution 
with a Lorentzian peak plus a polynomial background (the latter represented by the dotted line). The two vertical 
lines show the $3\Gamma$ cut applied to the data sample in order to select $\Sigma^-$ events. Only the fit of the 
data is shown.}
\label{fig:sig1_high}
\end{figure}

After the selection of $\gamma d \rightarrow K^+ \pi^- n (p)$ candidates, we looked for evidence of the 
presence of $\Sigma^-$ particles in the invariant-mass distribution of the pion and neutron. The 
distribution obtained for data and MC for the photon-energy bin 2.0--2.1 GeV is shown in 
Fig.~\ref{fig:sig1_high}. A sharp peak consistent with the $\Sigma^-$ appears on top of a small, almost 
flat background. Each distribution was fit with a Lorentzian peak plus a second-order polynomial for 
the background (in Fig.~\ref{fig:sig1_high} only the fit of the data is shown). The Lorentzian shape has been chosen 
because it reproduces the peak shape of both experimental 
and MC data better than the Gaussian. The final sample of $\gamma d \rightarrow K^+ \Sigma^- (p)$ events 
was obtained by selecting events within 3$\Gamma$ around the peak, where $\Gamma$ is the full width at half maximum 
of the Lorentzian. The background calculated by integrating the polynomial curve within the cuts was 
subtracted. The total background is generally increasing with the photon energy, and is between $2\%$ 
and $25\%$.

Finally, the extracted yield was corrected for the CLAS detector acceptance. For this, 
$\gamma d \rightarrow K^+ \Sigma^- p$ events were generated according to the Quark-Gluon Strings 
Model~\cite{QGSM,QGSMpc}. The Fermi motion of the neutron bound in the deuterium nucleus 
was described by the momentum distribution calculated from the Paris potential~\cite{paris}. The 
generated events were processed through a GEANT-based Monte Carlo simulation of the CLAS detector, 
incorporating all of the known subsystem efficiencies and resolutions. The simulated data were 
analyzed by the same software used in the real data processing and analysis. The CLAS acceptance was 
computed as the ratio between the number of events passing all the analysis cuts and the number of generated 
events in each one of the 100-MeV wide photon energy bins and 0.1-wide $\cos{\theta_K^{\rm{c.m.}}}$ bins.

The differential cross section for $K^+ \Sigma^-$ photoproduction on the neutron was calculated using 
the following relation:
\vspace{-0.7cm}

\begin{equation}
\frac{d\sigma}{d\Omega}= 
\frac {A}{\rho x N_A}
\frac{N^W_{peak}  b_{\Sigma^-}}{N_\gamma \, \Delta \Omega }
\left( 1 - B \right),
\label{eq:cross}
\end{equation}

\noindent
where $N^W_{peak}$ is the number of the \mbox{$\gamma d \rightarrow K^+ \Sigma^- (p) $} events weighted 
by the acceptance of the CLAS detector, $N_\gamma$ is the number of incident photons, $B$ is the fraction 
of background events, $A$ is the target molecular weight, $N_A$ is Avogadro's number, 
and $\rho=0.163$~g/cm$^3$ and $x=24$~cm are the target density and length, respectively. 
Photon absorption in the target was also calculated and found to be negligible. 
Systematic uncertainties of the final cross sections contain contributions from the photon flux 
calculation ($4\%$), target length and density ($0.5\%$), fiducial cuts ($1$--$3\%$, depending on the bin), 
background subtraction ($1$--$10\%$), neutron detection efficiency ($0.7\%$) and the Monte Carlo event 
generator ($1.7\%$). The total systematic uncertainty was obtained by adding in quadrature each 
contribution, bin by bin. Thus, the total systematic uncertainty in our cross section measurements is 
estimated to be about 4.5--13.5\%.

\begin{figure*}[ht]
\begin{center}
\vspace{4.cm}                       
\includegraphics{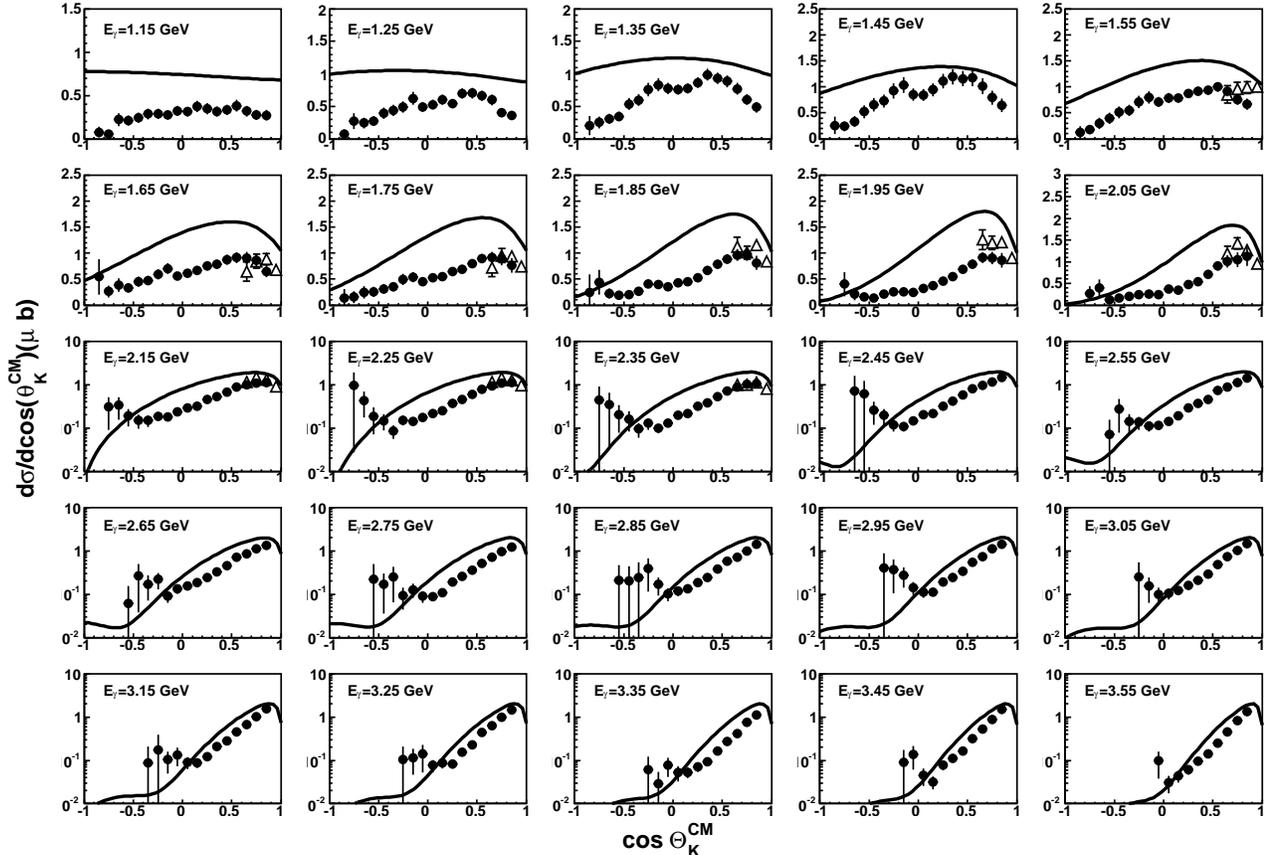}
\vspace{6.cm}
\caption[]{\small Differential cross sections of the reaction $\gamma d \rightarrow K^+ \Sigma^- (p)$ 
obtained by CLAS (full circles). The error bars represent the total (statistical plus systematic) uncertainty. 
LEPS data~\cite{LEPSn} (empty triangles) and a Regge-3 model 
prediction~\cite{model4} (solid curve) are also shown. Notice the logarithmic scale for high energy plots.}
\label{fig:xsc}
\end{center}
\end{figure*}

Our final results are shown as full circles in Fig.~\ref{fig:xsc}. 
For energies up to $E_{\gamma} =$ 2.1 GeV, the results are shown in linear scale while for 
higher energies, logarithmic scale has been chosen in order to make more readable the behavior at the backward angles. 
The error bars represent the total (statistical plus systematic) uncertainties. This is the first high-precision 
determination of $\Sigma^-$ photoproduction on the neutron covering a broad kaon-angle and photon-energy range. 
At a photon energy of $\sim 1.8$ GeV, a clear forward peak starts to appear and becomes more prominent 
as the photon energy increases. This behavior, that is typically attributed to contributions from 
$t$-channel mechanisms, is not observed at lower energies, where the dominant contributions appear to be 
from $s$-channel mechanisms. Above $\sim$2.1 GeV there are indications of a possible backward peak, 
which might suggest the presence of $u$-channel mechanisms.

The few LEPS data \cite{LEPSn} available for energies 1.5--2.4 GeV and at forward angles are shown in 
Fig.~\ref{fig:xsc}. Since these data have been provided in 50-MeV wide energy bins, for comparison with our 
results the weighted average of two bins has been computed and reported in the figure. They are in good 
agreement with our results within the total uncertainties.
 
Also shown in Fig.~\ref{fig:xsc} are the theoretical results of a Regge-based calculation 
(Regge-3 model)~\cite{model4}. In this model, the reaction amplitude incorporates the exchange of 
$K^+$ and $K^*(892)^+$ Regge trajectories. By adding resonance contributions to the Regge amplitudes, the 
model is able to describe the $\Lambda$ and $\Sigma^0$ photo- and electro-production data on the proton 
reasonably well~\cite{model1,model2,model3}. 
The Regge-based model overestimates our results at forward and intermediate angles by about a factor of two.  
At backward angles the calculated cross section is too small by an order or magnitude, which is a reflection 
of the lack of resonances in the model.

In conclusion, CLAS has provided the first precise determination of the $\gamma d \rightarrow K^+ \Sigma^- (p)$ 
cross section in a broad kinematic range where almost no data are available.
Since Final State Interaction (FSI) can be estimated to be small (less than 10\%) from calculations for the 
$\Lambda$ on the proton~\cite{guidal,laget}, the cross section on the free neutron are not expected to be significantly different. 
A comprehensive treatment of FSI and the extraction of the neutron cross section will be given in the 
forthcoming longer paper.
These results will significantly contribute to the improvement of the phenomenological analysis of meson 
photoproduction reactions at medium energies aiming to solve the missing resonance problem.

We would like to acknowledge the outstanding efforts of the staff of the Accelerator
and the Physics Divisions at JLab that made this experiment possible. 
This work was supported in part by the Italian Istituto Nazionale di Fisica Nucleare,
the French Centre National de la Recherche Scientifique and the Commissariat \`{a} l'Energie Atomique, 
the U.S. Department of Energy and the National Science Foundation, 
the National Research Foundation of Korea, and the UK Science and Technology Facilities Council
(STFC). The Southeastern Universities Research Association (SURA) operated the
Thomas Jefferson National Accelerator Facility for the United States
Department of Energy under contract DE-AC05-84ER40150 during this work.

\end{document}